# A NON-PERTURBATIVE, FINITE PARTICLE NUMBER APPROACH TO RELATIVISTIC SCATTERING THEORY[a]

Marcus Alfred[+], Petero Kwizera[#], James V. Lindesay[*,+], H. Pierre Noyes[*]

[*]Stanford Linear Accelerator Center, Stanford University, Stanford, California 94309
[+]Computational Physics Laboratory, Howard University, Washington, D.C. 20059
[#]Department of Physics, Makerere University, Kampala, Uganda

**Abstract**

We present integral equations for the scattering amplitudes of three scalar particles, using the Faddeev channel decomposition, which can be readily extended to any finite number of particles of any helicity. The solution of these equations, which have been demonstrated to be calculable, provide a non-perturbative way of obtaining relativistic scattering amplitudes for any finite number of particles that are Lorentz invariant, unitary, cluster decomposable and reduce unambiguously in the non-relativistic limit to the non-relativistic Faddeev equations. The aim of this program is to develop equations which explicitly depend upon physically observable input variables, and do not require "renormalization" or "dressing" of these parameters to connect them to the boundary states. As a unitary, cluster decomposible, multichannel theory, physical systems whose constituents are confined can be readily described.

---

[a] Work supported in part by Department of Energy contract DE-AC03-76SF00515

# I. Introduction

Despite the manifest success of perturbative relativistic quantum field theory in providing a rich and creative framework for describing elementary particle physics and exploring many quantitative predictions of QED, weak-electromagnetic unification, and QCD, the theory remains deeply flawed both at the foundational level and in its adequacy to provide a secure phenomenology for many important experimental situations.

One of these flaws was already noted in QED when it is renormalized by perturbative methods; Dyson[1] showed that the renormalized perturbation series is *not* uniformly convergent when extended beyond 137 terms, which convinced him that renormalized QED could never be a fundamental theory[2]. Another fundamental flaw is that there is no known way to derive a unique, static non-relativistic ``potential'' from *any* relativistic quantum field theory which can be used unambiguously in a non-relativistic quantum mechanical Hamiltonian. Among other things, this prevents the development of a rigorous model for low energy nuclear physics[3]. Basically, perturbative quantum field theory does not predict low lying bound states. This problem is further compounded in QCD; the basic mass parameters of the quarks cannot be directly measured because of confinement. Further, the mass parameters cannot be indirectly inferred from measurements of free and bound state masses below hadron production threshold because of ``infrared slavery'' (i.e., again the failure of perturbative methods). This paper offers an alternative approach, which could meet some of these challenges in a new way. The general attitude adopted in this paper is that basic theoretical structures should relate as closely as possible to the laboratory boundary conditions and detectable observables encountered in high energy particle physics. In particular the input mass parameters should be restricted to *physical* values.

The objective of the research program of which this paper forms a part is to formulate a non-perturbative, finite (though not fixed) particle number relativistic scattering theory which is Lorentz invariant, unitary, cluster decomposable and reduces unambiguously to a unique non-relativistic quantum dynamical formalism. Our program starts from the fact noted by Friedman, Lovelace and Namyslowski[4] and independently by Kowalski[5], that the unitarity of the non-relativistic three body amplitude defined by the Faddeev equations depends only on the unitarity of the two body input scattering amplitudes. The first paper in this program[6] provided a relativistic model of the Efimov effect for three scalar particles, which quantitatively predicted non-relativistic numerical results in the appropriate kinematic region that had been independently calculated by conventional means. The second[7] provided an explicit proof of the Lorentz invariance, unitarity and cluster decomposability of a formalism describing three scalar particles in a state of arbitrary angular momentum and the third[8] extended the formalism to include particle spin. The fourth[9] gave an explicit solution for the zero angular momentum states of three scalar particles driven by simple scalar unitary s-wave amplitudes. That model is self-consistent up to pair production threshold; however cluster decomposability was not investigated. Calculability was demonstrated in unpublished work using simplified models which allowed intermediate variables to be approximated[10]. Alfred[11] developed explicit general parameters for the intermediate states and explicitly demonstrated calculability using the formalism for a particular model. This development has removed many of the difficulties which have deterred the use of this formalism to describe physical systems. Applications of a related formalism which demonstrate the straightforward inclusion of properties such as confinement have been presented[12], and can be further developed using the formalism here presented. In this paper we exhibit the formalism in a way that explicitly meets *all* of the criteria listed in the first sentence of this paragraph. Papers extending the formalism to include anti-particles exhibiting the appropriate particle-antiparticle symmetries, including explicit models, and to describe quantum and pair creation, are in preparation.

Our motivation for developing this formalism includes the desire to meet a number of problems that any perturbative quantum theory of fields finds difficult or impossible to study, and to provide new and more powerful phenomenological tools in the few GeV region for data analysis at tau-charm factories, for nucleon form factor measurements, for Babar hadronic final state analysis, …….; these tools will also be useful for studying quark confinement at any energy, low or high. Further, perturbative and diagrammatic techniques are particularly difficult to apply in the theory of strong interactions, and in principle cannot be used as the starting point for many bound state calculations, since one cannot apply perturbative methods near singular points of an amplitude. In the past, scattering theoretic approaches have provided a means to incorporate both the scattering and bound state behaviors of a system within a single formalism, which was unitary (requiring no renormalization) and had physical parameters as inputs. However, since the full analytic structure of the system's behavior at all energies is manifest in scattering amplitudes, this approach has often been stifled by the sheer complexity which such complete solutions would entail. This complexity becomes magnified for relativistic scattering theory because of the nature of the energy-momentum dispersion relation, the expectation of particle-antiparticle production thresholds in the kinematics, and the difficulty to insure cluster decomposability in a covariant relativistic scattering formalism.

Faddeev[13] was successful in developing a formal mechanism for obtaining non-relativistic scattering amplitudes in a finite few particle number sector using inputs from the various scattering sub-clusters involved. His method uses a channel decomposition which recognizes all possible physical clusterings in the initial and final states, and guarantees unitarity and cluster decomposability of the physical amplitudes. The Faddeev channel decomposition allows one to eliminate singularities in kernels needed to calculate the scattering amplitudes due to disconnected diagrams. The formalism allows one to isolate the primary singularities due to bound states (and/or confined states) in the initial or final state, thus directly giving physical amplitudes from physical input parameters. This approach has advantages over many perturbative approaches in that there is no need for a renormalization procedure during or after the calculation (once the input amplitudes have been obtained); the input amplitudes contain the physical masses, charges, etc. of the interacting systems involved. However, because of the nature of the non-relativistic kinematics, one would require knowledge of the analytic structure of the input amplitudes off-shell far from physical values, and if those amplitudes have so called left-hand cuts the resulting equations could give ambiguous results[14].

In the mid 1980's, a scattering formalism was developed[7,8] which was shown to satisfy the following criteria in a finite fixed particle number sector:

1. it was fully Lorentz invariant, and on-shell generated full four-momentum conservation;

2. unitarity was assured for any scattering amplitude as long as the input interactions were unitary;

3. the amplitudes generated had unique fully off-shell forms, which were appropriate generalizations of the non-relativistic scattering formalism;

4. the formalism satisfied proper cluster decomposability;

5. all amplitudes and parameters involved in calculations have well defined non-relativistic limits.

The formalism utilized the channel decomposition of the scattering amplitudes developed by Faddeev, but was applicable for relativistic scattering systems. Amplitudes calculated using this formalism thus share all of the advantages of the Faddeev approach, allow the use of scattering theoretic techniques in relativistic regimes, and provide unique correspondence limits for non-relativistic few-particle scattering problems.

The development of several recent models[15] in nuclear and particle physics to describe the behavior of relativistic few particle systems demonstrate the need to expand understanding of the nature of these systems. Some approaches utilize techniques similar to the Bethe-Salpeter formalism[16], or Blankenbecler-Sugar approach[17] based on diagrammatic expansions, and quasi-potentials[18] in a relativistic model. Brayshaw[19] used a propagator that involved a linear difference in the four-momentum variables such that the form of the equations satisfied certain clustering properties. However, full cluster decomposability in the sense discussed here was not achieved due to complications arising from trying to simultaneously satisfy Lorentz kinematics and momentum conservation. Also popular are covariant formalisms that utilize light front dynamics[20] to define wave functions on a light front, which simplify the kernels needed for calculations. The approach developed here has the advantage of demonstrating the explicit analytic, unitarity, and clustering properties of the amplitudes, while maintaining the full relativistic and rotational invariance and one's intuitive connections to non-relativistic correspondence.

## II.  General Development of the Lorentz Invariant Scattering Equations

### A.  Covariant States

To begin, we exhibit the normalization conventions to be used in this development. Four-vectors will be denoted with arrows, and three-vectors will be under scripted, with the metric given by

$$\vec{A} \cdot \vec{B} = A^0 B^0 - \underline{A} \cdot \underline{B}$$

Four velocities will be normalized to unity, and are connected to the usual velocity ($\underline{\beta} = \frac{v}{c}$) by

$$\underline{u} \equiv \frac{\underline{\beta}}{(1-|\underline{\beta}|^2)^{\frac{1}{2}}} \text{ and } u^0 = (1+|\underline{u}|^2)^{\frac{1}{2}}$$

The states will transform through a unitary representation of the 10-dimensional Poincare' group $U(\Lambda,a)$ for Lorentz transformation $\Lambda$ and space-time translations $\vec{a}$:

$$(\Lambda_2, \vec{a}_2,) U(\Lambda_1, \vec{a}_1) = U(\Lambda_2 \Lambda_1, \vec{a}_2 + \Lambda_2 \vec{a}_1)$$

The standard state vector for a massive particle will be written as $\vec{k}_{(s)} = (m,0,0,0)$ whereas for a massless particle it will be written as $\vec{k}_{(s)} = (1,0,0,1)$ where the standard momenta have the appropriate units. Most of the development here will assume massive, scalar particles, although the general formalism including relativistic spin has been developed[8].

Single particle states will be normalized to satisfy

$$<\underline{k}'|\underline{k}> = \frac{\varepsilon(k)}{k^0_{(s)}} \delta^3(\underline{k}'-\underline{k})$$

where

$$\varepsilon(k) \equiv \sqrt{m^2 + |\underline{k}|^2}$$

with the completeness relations satisfying

$$1 = \int \frac{k^0_{(s)}}{\varepsilon(k)} d^3k \, |\underline{k}\rangle\langle\underline{k}| = \int d^4k \, |\vec{k}\rangle\langle\vec{k}| \, \delta(\sqrt{\vec{k}\cdot\vec{k}} - m).$$

This normalization is chosen, since amplitudes generated using these states will directly have the usual normalization of momentum states in non-relativistic scattering theory. It should be stated that the mass-shell constraint upon the interactions is not a precondition for this formulation. The scatterings need not preserve particle mass, and generally will not preserve sub-cluster invariant energy. These momentum basis states serve as the basis of most phenomenological data in terms of the asymptotic parameters of the particles. In this sense, the particles described are "fully dressed" and properly normalized such that the physically measured asymptotic parameters (masses, charges, etc.) are those represented in the states.

### B.    Two-particle scattering

In most of what follows we assume only pairwise interactions among the particles. This assumption does not restrict the generality of the method, since other interaction channels can be directly incorporated, but it simplifies demonstration of the method. For the development of a cluster decomposable formalism, it is more convenient to express the system parameters in terms of invariant energies rather than individual particle momenta. The relationship between the two-body variables is illustrated visually below:

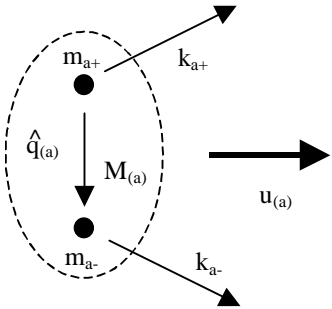

The channels will be denoted using the Faddeev notation in terms of the spectating cluster label. The various parameters are directly related through relativistic kinematics. The velocity $\underline{u}_{(a)}$ is used to boost from the two-particle center of momentum (2-CMS) frame to the arbitrary $k_{a+}$ and $k_{a-}$ momenta frame. The energies in the pair frame are given by,

$$\varepsilon_{a+}^{(a)} = \sqrt{m_{a+}^2 + \left|\underline{q}_{(a)}\right|^2} \quad \text{or} \quad \varepsilon_{a-}^{(a)} = \sqrt{m_{a-}^2 + \left|\underline{q}_{(a)}\right|^2}.$$

and the invariant pair energy $M_{(a)}$ is used to label the sum of pair energies in the pair rest frame:

$$M_{(a)} = \sqrt{m_{a+}^2 + \left|\underline{q}_{(a)}\right|^2} + \sqrt{m_{a-}^2 + \left|\underline{q}_{(a)}\right|^2} = \varepsilon_{a+}^{(a)} + \varepsilon_{a-}^{(a)}$$

2. 1

The invariant momentum of either particle in the center of momentum system satisfies

$$|\underline{q}_{(a)}(M_{(a)}^2, m_{a+}, m_{a-})|^2 \equiv \frac{[M_{(a)}^2 - (m_{a+} + m_{a-})^2][M_{(a)}^2 - (m_{a+} - m_{a-})^2]}{4M_{(a)}^2}$$

2. 2

A coordinate transformation can be used to relate the two equivalent representations of the parameter space:

$$\frac{m_{a+}}{\varepsilon_{a+}} d^3 k_{a+} \frac{m_{a-}}{\varepsilon_{a-}} d^3 k_{a-} \equiv \rho_{(a)}^{(2)}(M_{(a)}, m_{a+}, m_{a-}) dM_{(a)} \frac{d^3 u_{(a)}}{u_{(a)}^0} d^2 \hat{q}_{(a)}$$

2. 3

where

$$\rho_{(a)}^{(2)}(M_{(a)}, m_{a+}, m_{a-}) = m_{a+} m_{a-} M_{(a)}^2 | \underline{q}_{(a)}(M_{(a)}^2, m_{a+}, m_{a-}) |$$

2. 4

Particle scattering dynamics will be represented using the transition matrix $t_{(a)}$. Using our chosen parameterization, the amplitudes are defined by

$$\langle \underline{k}_{a+} \underline{k}_{a-} | t_{(a)}(\zeta) | \underline{k}_{a+0} \underline{k}_{a-0} \rangle \equiv (u_{(a)}^0) \delta^3(\underline{u}_{(a)} - \underline{u}'_{(a)}) \Theta(M_{(a)} - (m_{a+} + m_{a-})) \Theta(M_{(a)0} - (m_{a+} + m_{a-})) \otimes$$

$$\frac{\tau_{(a)}(M_{(a)}, \hat{q}_{(a)} | M_{(a)0}, \hat{q}_{(a)0}; \zeta)}{\sqrt{\rho_{(a)}^{(2)}(M_{(a)})} \sqrt{\rho_{(a)}^{(2)}(M_{(a)0})}}$$

2. 5

Henceforth the threshold theta functions will be understood and not exhibited explicitly, unless needed for clarity. The amplitudes $\tau_{(a)}$ satisfy off-shell (in the $\zeta$ parameter), off-diagonal (in the $M_{(a)}$ and $M'_{(a)}$ parameters) unitarity conditions given by

$$\tau_{(a)}(M_{(a)}, \hat{q}_{(a)} | M'_{(a)}, \hat{q}'_{(a)}; \zeta_1) - \tau_{(a)}(M_{(a)}, \hat{q}_{(a)} | M'_{(a)}, \hat{q}'_{(a)}; \zeta_2) =$$

$$(\zeta_2 - \zeta_1) \iint dM'' d^2\hat{q}'' \tau_{(a)}(M_{(a)}, \hat{q}_{(a)} | M'', \hat{q}''; \zeta_1) \left(\frac{1}{M'' - \zeta_1}\right)\left(\frac{1}{M'' - \zeta_2}\right) \tau_{(a)}(M'', \hat{q}'' | M'_{(a)}, \hat{q}'_{(a)}; \zeta_2)$$

2. 6

These amplitudes will be embedded in an interactive higher particle number space in the next section.

### III. **Few-Particle equations**

The dynamical (interacting) subcluster will be embedded in a higher particle number basis in a way that will preserve unitarity (total probability conservation), Lorentz invariance of amplitudes, and cluster decomposability. The total system will be decomposed in terms of a dynamical cluster and a spectating cluster. Either cluster can be considered to be a single particle, or a multitude of particles. The spectating cluster differs only in that its kinematics will enter parametrically with regards to the dynamical parameters of the dynamical cluster. Diagrammatically, this is represented as follows:

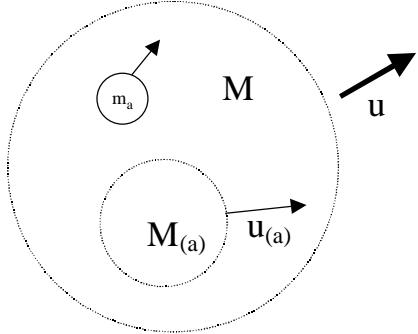

The particular channel is labeled using the Faddeev[13] convention, which labels the channel in terms of the spectating cluster. Clusters are defined to include all possible asymptotic configurations (including bound states) of the system. To demonstrate the formalism, we will restrict our considerations to three scalar particles that only interact pairwise, as mentioned in the previous section. One expects non-separable three-particle interactions in theories such as QCD, which has elementary three gluon vertices, and QED, where diagrams such as do not cluster decompose in the asymptotic regime. The inclusion of such interactions is straightforward by unitarily summing over all such non-separable terms and adding a three-particle Faddeev channel. However, such terms do not contribute to the primary singularities generated by pairwise bound initial or final states, and thus need not be included to demonstrate the method. The inclusion of particle spin is likewise straightforward, and can be accomplished using the well known angular momentum properties of the little group of transformations of the standard state vectors of the particles. One convenient method of including spin utilizes the helicity state representation of the particles due to Wick[21], and more generally developed in Jacob and Wick[22]. The helicity state representation was incorporated into our formalism by Markevich[8]. However, as previously mentioned, we feel the inclusion of particle spin would be an unnecessary detail in the discussion that follows, and we will restrict our presentation here to scalar particles.

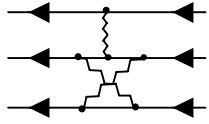

A. Boundary States

The most convenient parameterization of the kinematic variables is represented in the diagram that follows:

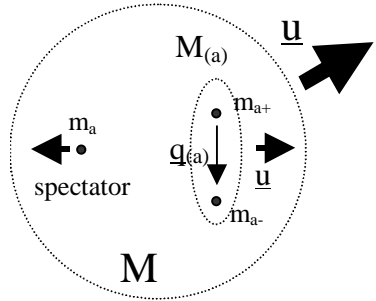

For this parameterization, M represents the invariant energy of the whole system, the three components $\underline{u}$ are the spatial components of the system's four-velocity, $m_a$ is the invariant energy of the spectating particle (or cluster), $M_{(a)}$ is the invariant energy of the dynamical pair with constituent masses $m_{a+}$ and $m_{a-}$, $\underline{u}_{(a)}$ are the spatial components of the four-velocity of the dynamical pair, and $\underline{q}_{(a)}$ is the invariant momentum (or the reversed momentum) of either particle in the pair center of momentum frame whose magnitude is fixed by $M_{(a)}$ using the formula [2. 2]. As indicated in the diagram, our convention requires the unit vector $\hat{q}_{(a)}$ in the direction which points from $m_{a+}$ to $m_{a-}$. The parameters are related using Lorentz transformations between the pair and three-particle center of momentum systems. The pair momenta define the invariant pair energy and 4-velocity

$$M_{(a)}\vec{u}_{(a)} = \vec{k}_{a+} + \vec{k}_{a-}$$

and the momentum of either particle in the center of momentum pair is related to its general momentum by a Lorentz transformation

$$k_{a-}^{\mu} = \Lambda^{\mu}{}_{\nu}(\underline{u}_{(a)})k_{a-}^{*\,\nu}$$

$$\vec{k}_{a-}^{*} = \left(\sqrt{m_{a-}^2 + q_{(a)}^2}, |\underline{q}_{(a)}(M_{(a)}, m_{a+}, m_{a-})|\hat{q}_{(a)}\right)$$

where the magnitude of $\underline{q}_{(a)}$ has been previously expressed. The three-particle four-momentum satisfies

$$M\vec{u} = \vec{k}_a + M_{(a)}\vec{u}_{(a)}$$

**3. 1**

which can be solved for the invariant energy M

$$M = \sqrt{m_a^2 + M_{(a)}^2 - 2M_{(a)}(\vec{k}_a \cdot \vec{u}_{(a)})}$$

**3. 2**

These kinematic relationships will allow a direct change of variables from the particle momenta $\{\underline{k}_a\}$ to the more convenient invariant energies and four velocities. The invariant factor is defined using the following identifications

$$\frac{k_{1(s)}^0}{\varepsilon_1}d^3k_1 \frac{k_{2(s)}^0}{\varepsilon_2}d^3k_2 \frac{k_{3(s)}^0}{\varepsilon_3}d^3k_3 = \rho_{(a)}^{(3)}(M, u_{(a)}^0)dM\frac{d^3u}{u^0}\frac{d^3u_{(a)}}{u_{(a)}^0}d^2\hat{q}_{(a)}$$

**3. 3**

and can be calculated using

$$\rho_{(a)}^{(3)}(M, u_{(a)}^0) = \frac{u^0 u_{(a)}^0 k_{1(s)}^0 k_{2(s)}^0 k_{3(s)}^0}{\varepsilon_1 \varepsilon_2 \varepsilon_3} \frac{\partial(\underline{k}_a, \underline{k}_{a+}, \underline{k}_{a-})}{\partial(\underline{k}_a, \underline{u}_{(a)}, \underline{k}_{a-})} \frac{\partial(\underline{k}_a, \underline{u}_{(a)}, \underline{k}_{a-})}{\partial(\underline{k}_a, \underline{u}_{(a)}, \underline{q}_{(a)})} \otimes$$

$$\frac{\partial(\underline{k}_a, \underline{u}_{(a)}, \underline{q}_{(a)})}{\partial(\underline{k}_a, \underline{u}_{(a)}, |\underline{q}_{(a)}|, \hat{q}_{(a)})} \frac{\partial(\underline{k}_a, \underline{u}_{(a)}, |\underline{q}_{(a)}|, \hat{q}_{(a)})}{\partial(M, \underline{k}_a, \underline{u}_{(a)}, \hat{q}_{(a)})} \frac{\partial(M, \underline{k}_a, \underline{u}_{(a)}, \hat{q}_{(a)})}{\partial(M, \underline{u}, \underline{u}_{(a)}, \hat{q}_{(a)})}$$

This allows different equivalent parametric descriptions of the kinematics of the boundary states. The form of the Jacobian factor $\rho_{(a)}^{(3)}$ is given by[11]

$$\rho^{(3)}_{(a)}(M, u^0_{(a)}) = \frac{k^0_{(s)a} k^0_{(s)a+} k^0_{(s)a-} M^3 M^2_{(a)} \mid \underline{q}_{(a)}(M^2_{(a)}, m_{a+}, m_{a-}) \mid}{M u^0_{(a)} - M_{(a)}}$$

**3. 4**

where the invariant dynamical cluster energy is given by

$$M_{(a)} = M u^0_{(a)} - \sqrt{m^2_a + M^2 u^2_{(a)}}$$

**3. 5**

As explained in our discussion of normalization conventions in Section II.A, the factors $k_{(s)}^0$ represents the zeroth component of the standard state vector, conventionally given by the mass m for massive particles, and 1 for massless particles.

The physical problem will be described in terms of the boundary states that satisfy the asymptotic conditions. The possible asymptotic situations consist of bound pairs with a third (only self-interacting) particle, three free (only self interacting) particles, or perhaps a three-particle bound state, which will occur as a singularity in the off-shell parameter. The boundary states will be represented using

$\underline{k_1}, m_1$ ⟵
$\underline{k_2}, m_2$ ⟵
$\underline{k_3}, m_3$ ⟵   $\left| \Phi_{(0)} : (\underline{k}_1, m_1; \underline{k}_2, m_2; \underline{k}_3, m_3) \right\rangle$ for the three particle asymptotic states, and

$\underline{k}_a, m_a$ ⟵
$\underline{p}_{(a)}, \mu_a$ ⟵   $\left| \Phi_{(a)} : \underline{k}_a, m_a; \psi_{(a)}(\underline{p}_{(a)}, \mu_{(a)}, l_{(a)}, l_{z(a)}) \right\rangle$ for bound pair + spectator,

where the bound pair has mass $\mu_{(a)}$ and momentum $\underline{p}_{(a)}$. Of course these states can be alternatively written in terms of momentum variables or angular momentum states (usually the bound state will not be represented using the two continuous degrees of freedom represented by $\hat{q}_{(a)}$ but instead using the angular momentum eigenvalues $l$ and $l_z$). These boundary states are presumed to be eigenstates of "asymptotic" Hamiltonians with additive energies for the particles.

The bound state wave functions are assumed to satisfy the same normalization conditions as any non-composite particle

$$< \Phi_{(a)} : k'_a\, m_a; \psi_{(a)}(\underline{p}'_{(a)}, \mu^r_{(a)}, l_{(a)}', l_{z(a)}') | \Phi_{(a)} : k_a m_a; \psi_{(a)}(\underline{p}_{(a)}, \mu^s_{(a)}, l_{(a)}, l_{z(a)}) >=$$

$$\delta_{rs} \delta_{l'_{(a)} l_{(a)}} \delta_{l'_{z(a)} l_{z(a)}} \frac{\varepsilon_{\mu^r_{(a)}}}{\mu^r_{(a)}} \delta^3(\underline{p}'_{(a)} - \underline{p}_{(a)}) \frac{\varepsilon_a}{m_a} \delta^3(\underline{k}_a' - \underline{k}_a) =$$

$$\delta_{rs} \delta_{l'_{(a)} l_{(a)}} \delta_{l'_{z(a)} l_{z(a)}} \frac{\sqrt{\mu^{r\,2}_{(a)} + \mu^{r\,2}_{(a)} u^2_{(a)}}}{\mu^r_{(a)}} \delta^3(\mu^r_{(a)} \underline{u}'_{(a)} - \mu^s_{(a)} \underline{u}_{(a)}) \frac{\varepsilon_a}{m_a} \delta^3(\underline{k}_a' - \underline{k}_a)$$

3. 6

To represent these energy eigenstates in a momentum basis involves the usual kinematic factors:

$$< \underline{k}'_{a+}\, \underline{k}'_{a-} | \psi_{(a)}(\underline{p}_{(a)}, \mu_{(a)}, l_{(a)}, l_{z(a)}) >\equiv u^0_{(a)} \delta^3(\underline{u}'_{(a)} - \frac{\underline{p}_{(a)}}{\mu_{(a)}}) \frac{\psi_{(a)}(M'_{(a)}; \mu_{(a)}, l_{(a)}, l_{z(a)}) Y^{l_{z(a)}}_{l_{(a)}}(\hat{q}'_{(a)})}{[\rho^{(2)}_{(a)}(M'_{(a)}, m_{a+}, m_{a-})]^{\frac{1}{2}}}$$

$$= \mu^3_{(a)} \frac{\varepsilon_{\mu_{(a)}}}{\mu_{(a)}} \delta^3(\mu_{(a)} \underline{u}'_{(a)} - \underline{p}_{(a)}) \frac{\psi_{(a)}(M'_{(a)}; \mu_{(a)}, l_{(a)}, l_{z(a)}) Y^{l_{z(a)}}_{l_{(a)}}(\hat{q}'_{(a)})}{[\rho^{(2)}_{(a)}(M'_{(a)}, m_{a+}, m_{a-})]^{\frac{1}{2}}}$$

3. 7

which normalizes the bound state wave functions as follows:

$$\int dM'_{(a)} \left| \psi_{(a)}(M'_{(a)}; \mu^s_{(a)}, l_{(a)}, l_{z(a)}) \right|^2 = 1$$

3. 8

Notice that $\mu_{(a)} \underline{u}'_{(a)} \neq \underline{k}'_{a+} + \underline{k}'_{a-} = M'_{(a)} \underline{u}'_{(a)}$ except in the pair zero-momentum frame. The three particle boundary state constructed from this pair state will similarly be presumed to be an eigenstate of an asymptotic Hamiltonian with additive energies for the discrete pair + third particle system

$$H_{(a)} \left| \Phi_{(a)} : \underline{k}_a, m_a; \psi_{(a)}(\underline{p}_{(a)}, \mu_{(a)}, l_{(a)}, l_{z(a)}) \right\rangle = (\varepsilon_a + \varepsilon_{\mu_{(a)}}) \left| \Phi_{(a)} : \underline{k}_a, m_a; \psi_{(a)}(\underline{p}_{(a)}, \mu_{(a)}, l_{(a)}, l_{z(a)}) \right\rangle$$

$$\varepsilon_a = \sqrt{m_a^2 + k_a^2}$$

$$\varepsilon_{\mu_{(a)}} = \sqrt{\mu^2_{(a)} + p^2_{(a)}}$$

These pair bound states will produce primary singularities in the three-particle amplitudes when those amplitudes are expressed in terms of the three "free" momentum states, as will be demonstrated later when extracting the physical observables.

### B. Scattering operators

The three-particle transition operator $T(Z)$ will be decomposed using the Faddeev method:

$$T(Z) = \sum_{ab} T_{ab}(Z)$$

**3. 9**

where the components $T_{ab}$ satisfy

$$T_{ab}(Z) = \delta_{ab} T_{(a)}(Z) - \sum_{d} \overline{\delta}_{ad} T_{(a)}(Z) R_{o}(Z) T_{db}(Z)$$

$$\overline{\delta}_{ad} \equiv 1 - \delta_{ad}$$

**3. 10**

The amplitudes $T_{(a)}(Z)$ represent the two-particle input amplitudes embedded in the three-particle space, and the resolvant is given by

$$R_{o}(Z) = \frac{1}{H_{o} - Z}$$

**3. 11**

Cluster decomposability requires that this resolvant be linear in the energies which add asymptotically.

The equation for the three-particle scattering amplitudes requires input from two-particle scatterings. In previous work[7] we required that the form of the input amplitudes $T_{(a)}(Z)$ be restricted in a way that preserved Lorentz invariance and cluster decomposability. It was found that this could be achieved by keeping the interacting cluster and the total system in the same velocity coordinate systems. This led to the form:

$$\langle \underline{k}_a \underline{k}_{a+} \underline{k}_{a-} | T_{(a)}(Z) | \underline{k}_a' \underline{k}_{a+}' \underline{k}_{a-}' \rangle = (u^0) \delta^3(\underline{u} - \underline{u}')(u_{(a)}^0) \delta^3(\underline{u}_{(a)} - \underline{u}_{(a)}') \frac{\tau_{(a)}(\omega_{(a)}, \hat{q}_{(a)} | \omega_{(a)}', \hat{q}_{(a)}'; \zeta_{(a)})}{\sqrt{[\rho_{(a)}^{(3)}(M, u_{(a)}^0) \rho_{(a)}^{(3)}(M', u_{(a)}^0)]}}$$

**3. 12**

where the Jacobian factors are given in equation 3. 4. The off-shell and off-diagonal behavior of this amplitude is of a special form if we are to ensure its cluster decomposability. The parameter $\omega_{(a)}$ replaces the off-diagonal energy in the two-particle amplitude, and is given by

$$\omega_{(a)} = \frac{M - e_a}{\vec{u} \cdot \vec{u}_{(a)}}$$

**3. 13**

The parameter $\zeta_{(a)}$ replaces the off-shell energy parameter $Z$:

$$\zeta_{(a)} = \frac{Z - e_a}{\vec{u} \cdot \vec{u}_{(a)}}$$

3. 14

Here, the parameteric energy $e_a$ represents the (on-shell) energy of the spectating cluster and properly includes the spectator's kinematics in the total system. When expressed in the 3-particle center of momentum system, we find that

$$\vec{u} \cdot \vec{u}_{(a)} \to u^0_{(a)} = \sqrt{1 + u^2_{(a)}}$$
$$M_o \equiv \Re e[Z_{OnShell}]$$
$$e_a(M_o, u^0_{(a)}) \equiv u^0_{(a)} \sqrt{m_a^2 + M_o^2 u^2_{(a)}} - M_o u^2_{(a)}$$

3. 15

If asymptotically the dynamical cluster is a bound state of invariant energy $\mu_{(a)}$, the spectating cluster parameter can equivalently be written as

$$e_a(M_o, \mu_{(a)}) = \frac{1}{2M_o}(M_o^2 + m_a^2 - \mu^2_{(a)})$$

The invariant energy off-shell and off-diagonal parameters are directly related to on-shell values of the invariant energy of the pair

$$\omega_{(a)} \xrightarrow{on-diagonal} M_{(a)}$$
$$\zeta_{(a)} \xrightarrow{on-shell} M_{(a)o}$$ ,

but the form of these parameters is crucial to the clustering properties of the equations, as will be discussed later. The non-relativistic limit of the cluster decomposable amplitude will be directly related to the non-relativistic Faddeev amplitude in Section IV.B.

Diagrammatically the equation for $T_{ab}$ can be represented as follows:

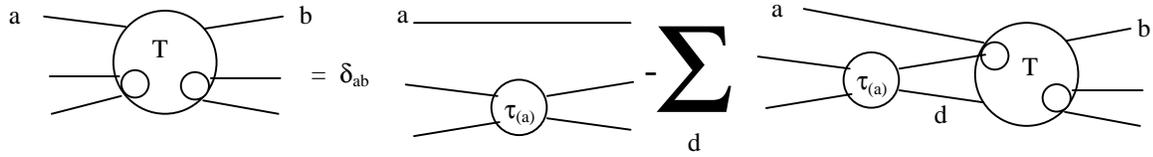

The disconnected term will generate a singular kernel; we define the fully connected amplitude by:

$$W_{ab}(Z) \equiv T_{ab}(Z) - \delta_{ab} T_{(a)}(Z)$$

**3. 16**

It is generated by well defined, non-singular kernels. The full resolvant can be expressed either using the full scattering amplitude *T(Z)* by:

$$R_F(Z) = R_o(Z) - R_o(Z) \sum_{ab} T_{ab}(Z) R_o(Z)$$

**3. 17**

or in terms of the various channel resolvants by:

$$R_F(Z) = R_o(Z) + \sum_a [R_a(Z) - R_o(Z)] - R_o(Z) \sum_{ab} W_{ab}(Z) R_o(Z)$$

**3. 18**

We use these equations to generate the physical scattering amplitudes in the next section.

The connected amplitudes $W_{ab}$ satisfy the equation formally expressed by:

$$W_{ab}(Z) = -\bar{\delta}_{ab} T_{(a)}(Z) R_o(Z) T_{(b)}(Z) - \sum_d \bar{\delta}_{ad} T_{(a)}(Z) R_o(Z) W_{db}(Z)$$

**3. 19**

which can be represented by the following diagram:

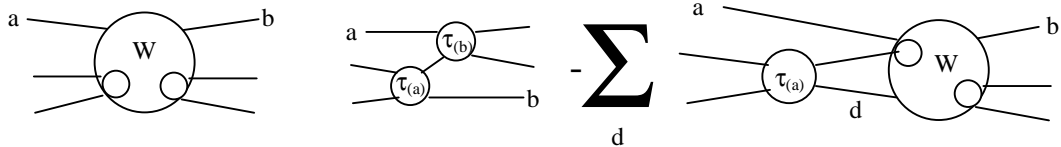

Consistent with the parameterization needed to ensure cluster decomposability, the connected amplitude is defined by

$$\langle \underline{k}_1 \underline{k}_2 \underline{k}_3 | W_{ad}(Z) | \underline{k}'_1 \underline{k}'_2 \underline{k}'_3 \rangle = (u^0) \delta^3(\underline{u} - \underline{u}') \frac{W_{ad}(M, \underline{u}_{(a)}, \hat{q}_{(a)} | M', \underline{u}'_{(d)}, \hat{q}'_{(d)}; Z)}{[\rho^{(3)}_{(a)}(M, u^0_{(a)}) \rho^{(3)}_{(d)}(M', u'^0_{(d)})]^{\frac{1}{2}}}$$

3. 20

where the pair velocities $\underline{u}_{(a)}$ henceforth will be assumed to be expressed relative to the three-particle center of momentum system $\underline{u}=\underline{0}$. If a complete set of momentum states are used in the sum, the formal expression 3. 19 becomes an integral equation. Since the two-particle input amplitudes given by equation 3. 12 contain a Lorentz-frame-conserving delta function, there will only be three degrees of freedom in the intermediate integrals, i.e. three of the six variables $M', \underline{u}'_{(d)}, \hat{q}'_{(d)}$ are constrained by the intermediate state kinematics:

$$W_{ab}(M, \underline{u}_{(a)}, \hat{q}_{(a)} | M_0, \underline{u}_{0(b)}, \hat{q}_{0(b)}; Z) = W^D_{ab}(M, \underline{u}_{(a)}, \hat{q}_{(a)} | M_0, \underline{u}_{0(b)}, \hat{q}_{0(b)}; Z) +$$

$$- \sum_d \bar{\delta}_{ad} \int \frac{d^3 u'_{(d)}}{u^{0'}_{(d)}} \tau_{(a)}(\omega_{(a)} \hat{q}_{(a)} | \omega'_{(a)} \hat{q}'_{(a)}; \zeta_{(a)}) F_{(ad)}(u_{(a)}, u'_{(d)}, \hat{u}_{(a)} \cdot \hat{u}'_{(d)}) \otimes$$

$$\frac{1}{M'-Z} W_{db}(M', \underline{u}'_{(d)}, \hat{q}'_{(d)} | M_0, \underline{u}_{0(b)}, \hat{q}_{0(b)}; Z)$$

3. 21

where the function $F_{(ad)}$ contains the Jacobians and kinematic constraints coming from the change of variables:

$$F_{(ad)}(u_{(a)}, u'_{(d)}, \hat{u}_{(a)} \cdot \hat{u}'_{(d)}) \equiv \int \frac{\rho^{(3)}_{(d)}(M', u^0_{(d)}) dM' d^2 \hat{q}'_{(d)} u^0_{(a)} \delta^3(\underline{u}_{(a)} - \underline{u}'_{(a)})}{[\rho^{(3)}_{(a)}(M', u^0_{(a)}) \rho^{(3)}_{(d)}(M', u^0_{(d)})]^{\frac{1}{2}}}$$

3. 22

The driving term for the integral equation, which can only have primary singularities due to any two-particle bound states that might appear in the $\zeta_{(a)}$ parameters, is given by

$$W^D_{ab}(M, \underline{u}_{(a)}, \hat{q}_{(a)} | M_o, \underline{u}_{(b)o}, \hat{q}_{(b)o}; Z) =$$

$$-\bar{\delta}_{ab} \tau_{(a)}(\omega_{(a)}, \hat{q}_{(a)} | \omega'_{(a)}, \hat{q}'_{(a)}; \zeta_{(a)}) \frac{F_{(ab)}(u_{(a)}, u_{(b)o}, \hat{u}_{(a)} \cdot \hat{u}_{(b)o})}{M' - Z} \tau_{(b)}(\omega'_{(b)}, \hat{q}'_{(b)} | \omega_{0(b)}, \hat{q}_{0(b)}; \zeta_{(b)})$$

3. 23

where the primed parameters are defined by kinematic consistency coming from integrating over the two Lorentz frame conserving delta functions $u_{(a)}^0 \delta^3(\underline{u}_{(a)} - \underline{u}_{(a)}')$ and $u_{(b)o}^0 \delta^3(\underline{u}_{(b)o} - \underline{u}_{(b)}')$ in the three-particle center of momentum system. This form of scattering equations is seen to satisfy Lorentz invariance and angular momentum conservation. Since the resolvant will generate a diagonalizing invariant energy delta function, this form will give energy and momentum conservation on-shell from the overall Lorentz velocity frame conservation. The cluster decomposability of these amplitudes will be discussed later.

It is interesting to note the off-diagonal behavior of the particle a in the intermediate state at this point. Examine the kinematics in the region along the dashed line in the following diagram:

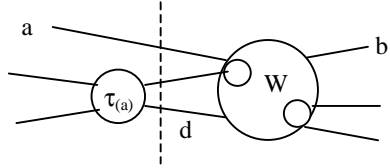

The four-momentum of particle a in the off-diagonal state using our parameterization of the two-particle embedding will be represented by $\vec{\kappa}_a$, and it has the form

$$\vec{\kappa}_a = \left(e_a(M_o, u^0_{(a)}), -\omega_{(a)}(M', u^0_{(a)})\underline{u}_{(a)}\right).$$

By direct algebra, it is seen that the invariant energy of this particle is given by

$$\vec{\kappa}_a \cdot \vec{\kappa}_a = m_a^2 + [M_o(M_o - 2e_a) - M'(M'-2e_a)]\left(\frac{|\underline{u}_{(a)}|}{u^0_{(a)}}\right)^2$$

Note that particle a goes off mass-shell in a manner similar to what occurs in perturbative relativistic quantum field theories (like QED); the integration variable M' could alternatively be expressed in terms of integrals over the parameter $\vec{\kappa}_a \cdot \vec{\kappa}_a$ or $d^4\kappa_a$. However the clustering properties are not apparent using that parameterization.

In prior published work[7] the form of the function $F_{(ad)}$ was only expressed implicitly in terms of the kinematic parameters. However, by following the chain of transformations

$$\frac{\partial(M', \hat{q}'_{(d)})}{\partial(\underline{u}'_{(a)})} = \frac{\partial(M', \hat{q}'_{(d)})}{\partial(M_{(d)}, \hat{q}'_{(d)})} \frac{\partial(M_{(d)}, \hat{q}'_{(d)})}{\partial(\underline{q}_{(d)})} \frac{\partial(\underline{q}_{(d)})}{\partial(\underline{u}'_{(a)})}$$

Alfred[11] was able to explicitly derive a form for the function using this Jacobian. A Lorentz transformation is necessary to transform kinematic parameters in the (a) channel to those in the (d) channel, shown diagrammatically as follows:

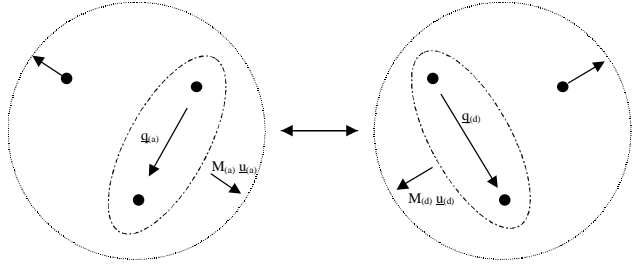

Using the Lorentz transformations in the form given by

$$L^0_{\ 0}(\underline{u}) = u^0$$
$$L^0_{\ k}(\underline{u}) = -u_k$$
$$L^j_{\ 0}(\underline{u}) = -u_j$$
$$L^j_{\ k}(\underline{u}) = \delta_{jk} + \frac{u_j u_k}{u^0 + 1}$$

one can directly relate kinematic variables in the two different coordinate systems

$$M'_{(a)} \underline{u}'_{(a)} = \left[ -\varepsilon_a^{(d)} + \frac{\underline{u}'_{(d)} \cdot \underline{q}'_{(d)}}{u'^0_{(d)} + 1} \right] \underline{u}'_{(d)} + \underline{q}'_{(d)}$$

$$M'_{(a)} u'^0_{(a)} = \varepsilon_d + u'^0_{(d)} \varepsilon_{d-}^{(d)} + \underline{u}'_{(d)} \cdot \underline{q}'_{(a)}$$

$$M'^2_{(a)} = \left( M' u'^0_{(d)} - \varepsilon_a^{(d)} \right)^2 - | M' \underline{u}'_{(d)} - \underline{q}'_{(a)} |^2$$

$$M' = M'_{(d)} u'^0_{(d)} + \varepsilon_d$$
$$M'_{(d)} = \varepsilon_a^{(d)} + \varepsilon_{d-}^{(d)}$$

**3. 24**

where $\underline{q}'_{(d)}$ is the momentum (or reversed momentum) of either particle in the pair (d) center of momentum frame (see figure in Section III.A.), and $\varepsilon_a^{(d)}$ is the energy of the particle a (which is one of the pair a and d-) in that frame as expressed in equation 2. 1. This allows the form of the factor $F_{(ad)}$ to be explicitly calculated:

$$F_{(ad)}\left(\underline{u}_{(a)}, \underline{u}'_{(d)}, \hat{u}_{(a)} \cdot \hat{u}'_{(d)}\right) = \left[\frac{\varepsilon_a^{(a)} \varepsilon_d^{(d)}}{q'_{(a)} q'_{(d)}}\right]^{\frac{1}{2}} \frac{M'^2_{(a)} M'^2_{(d)}}{\varepsilon'_d \varepsilon_a^{(d)} \varepsilon_{d-}^{(d)} u^0_{(a)}} \otimes$$

$$\{1 + \underline{\alpha}_{(ad)} \cdot \underline{u}_{(a)} + \underline{\beta}_{(ad)} \cdot \underline{u}'_{(d)} + (\underline{\alpha}_{(ad)} \cdot \underline{\beta}_{(ad)})(\underline{u}_{(a)} \cdot \underline{u}'_{(d)}) - (\underline{\alpha}_{(ad)} \cdot \underline{u}'_{(d)})(\underline{\beta}_{(ad)} \cdot \underline{u}_{(a)})\}^{-1}$$

where

$$\varepsilon_a^{(a)} \equiv M' u^0_{(a)} - M'_{(a)}$$

$$\varepsilon_a^{(a)} \equiv M' u'^0_{(d)} - M'_{(d)}$$

$$\underline{\alpha}_{(ad)}(\underline{u}_{(a)}, \underline{u}'_{(d)}) \equiv -\underline{\nabla}_{(d)} M'_{(a)}$$

$$\underline{\beta}_{(ad)}(\underline{u}_{(a)}, \underline{u}'_{(d)}) \equiv \frac{\underline{u}'_{(d)}}{u'^0_{(d)} + 1} - \underline{\nabla}_{(d)} \varepsilon_a^{(d)}$$

$$\left(\underline{\nabla}_{(d)}\right)_j \equiv \frac{\partial}{\partial q'_{(d)j}}$$

3. 25

and the relationship between the magnitude of the momentum q'$_{(d)}$ and the invariant energy M'$_{(d)}$ is given by equation 2. 1 and equation 2. 2 for mass-shell boundary state expansion of intermediate states.

For completeness, the forms of the energy parameters which are derivable from relativistic kinematics, will be exhibited below:

$$\underline{\nabla}_{(d)} M'_{(a)} = \left[\left(M' - u'^0_{(d)} \varepsilon_a^{(d)} + \underline{u}'_{(d)} \cdot \underline{q}'_{(d)}\right) \underline{\nabla}_{(d)} M' - \left(M' u'^0_{(d)} - \varepsilon_a^{(d)}\right) \underline{\nabla}_{(d)} \varepsilon_a^{(d)} + M' \underline{u}'_{(d)} - \underline{q}'_{(d)}\right] / M'_{(a)}$$

$$\underline{\nabla}_{(d)} M' = \left(u'^0_{(d)} + \frac{M'_{(d)} u'^2_{(d)}}{\varepsilon'_d}\right) \frac{M'_{(d)} \underline{q}'_{(d)}}{\varepsilon_a^{(d)} \varepsilon_{d-}^{(d)}}$$

$$\underline{\nabla}_{(d)} \varepsilon_a^{(d)} = \frac{\underline{q}'_{(d)}}{\varepsilon_a^{(d)}}$$

$$\varepsilon_d' = \sqrt{m_d^2 + M'^2_{(d)} u'^2_{(d)}}$$

3. 26

The total invariant energy M' can then be expressed by solving the (complicated) kinematic consistency equations using

$$M' = M'_{(a)} u^0_{(a)} + \sqrt{m_a^2 + M'^2_{(a)} u^2_{(a)}} = M'_{(d)} u'^0_{(d)} + \sqrt{m_d^2 + M'^2_{(d)} u'^2_{(d)}}$$

3. 27

Thus, the equations for all parameters in the integral equation for the scattering amplitudes $W_{ab}$ have been explicitly demonstrated.

## Angular momentum decomposition

The form of the equation 3.21 for the scattering amplitudes involves a three-dimensional integral equation over intermediate state kinematic parameters. It is often convenient to reduce the dimensionality of the integration by doing an angular momentum decomposition of the amplitudes. Since the angular momentum space $|l,l_z\rangle$ will be assumed to be a complete 2-dimensional basis, only an integration over one continuous parameter will remain after this decomposition. Since overall angular momentum conservation is preserved, we will decompose the overall amplitude as follows:

$$W_{ab}\left(M, \underline{u}_{(a)}, \hat{q}_{(a)} \mid M_o, \underline{u}_{(b)o}, \hat{q}_{(b)o}; Z\right) =$$
$$\sum_{\substack{L, l_a l_{az}, \\ l_{bo} l_{bzo}}} Y_{l_a}^{l_{az}}(\hat{q}_{(a)}) W_{ab}^L\left(u_{(a)}, l_a l_{az} \mid u_{(b)o}, l_{bo} l_{bzo}; Z\right) Y_{l_{bo}}^{l_{bzo}*}(\hat{q}_{(b)o}) \left(\frac{2L+1}{4\pi}\right) P_L(\hat{u}_{(a)} \cdot \hat{u}_{(b)o})$$

3. 28

The two-particle input amplitudes are similarly decomposed into appropriate partial waves

$$\tau_{(a)}(\omega_{(a)}, \hat{q}_{(a)} \mid \omega'_{(a)}, \hat{q}'_{(a)}; \zeta_{(a)}) = \sum_{l_{(a)} l_{z(a)}} Y_{l_{(a)}}^{l_{z(a)}}(\hat{q}_{(a)}) \tau_{(a)}^{\ell_a}\left(\omega_{(a)} \mid \omega'_{(a)}; \zeta_{(a)}\right) Y_{l_{(a)}}^{l_{z(a)}*}(\hat{q}'_{(a)})$$

3. 29

The integral equation for a system with total (three-particle) angular momentum L is then given by

$$W_{ab}^L\left(u_{(a)}, \ell_{(a)}, \ell_{z(a)} \mid u_{(b)o}, \ell_{(b)o}, \ell_{z(b)o}; Z\right) = W_{ab}^{(D)L}\left(u_{(a)}, \ell_{(a)}, \ell_{z(a)} \mid u_{(b)o}, \ell_{(b)o}, \ell_{z(b)o}; Z\right) +$$
$$- \sum_d \bar{\delta}_{ad} \sum_{\ell'_{(d)}, \ell'_{z(d)}} \int du'_{(d)} \, K_{ad}^L\left(u_{(a)}, \ell_{(a)}, \ell_{z(a)} \mid u'_{(d)}, \ell'_{(d)}, \ell'_{z(d)}; Z\right) W_{db}^L\left(u'_{(d)}, \ell'_{(d)}, \ell'_{z(d)} \mid u_{(b)o}, \ell_{(b)o}, \ell_{z(b)o}; Z\right)$$

3. 30

where the components of the kernel are defined by

$$K_{ad}^{L}\left(u_{(a)}, \ell_{(a)}, \ell_{z(a)} \mid u'_{(d)}, \ell'_{(d)}, \ell'_{z(d)}; Z\right) = -\overline{\delta}_{ad} \frac{u'^{2}_{(d)}}{u^{0'}_{(d)}} \frac{1}{2\pi} \otimes$$

$$\int_{-1}^{1} d\left(\hat{u}_{(a)} \cdot \hat{u}'_{(d)}\right) \tau_{(a)}^{\ell_{(a)}}\left(\omega_{(a)} \mid \omega'_{(a)}; \zeta_{(a)}\right) Y_{\ell_{(a)}}^{\ell_{z(a)}*}\left(\hat{q}'_{(a)}\right) \frac{F_{(ad)}\left(u_{(a)}, u'_{(d)}, \hat{u}_{(a)} \cdot \hat{u}'_{(d)}\right)}{M' - Z} Y_{\ell'_{(d)}}^{\ell'_{z(d)}}\left(\hat{q}'_{(d)}\right) P_{L}\left(\hat{u}_{(a)} \cdot \hat{u}'_{(d)}\right)$$

3. 31

and the invariant energy M' is defined consistent with the relativistic kinematics of the intermediate state in equation 3. 27 . The angular momentum decomposition is seen to reduce the equation into the form of a one -dimensional multi-channel integral equation. A more complete discussion of the angular momentum properties can be found in the references[8,21].

### C. Unitarity

The unitarity of the fully off-shell (off-shell and off-diagonal) Faddeev equation follows from the unitarity of the two-particle input amplitudes as was shown by Freedman, Lovelace, and Namyslowski[4], and independently by Kowalski[5]. If the two-particle amplitudes have been properly embedded in the three-particle space for a relativistic theory, the form of the proof remains essentially unchanged. One first notes that the embedded input amplitude satisfies a unitarity condition of the form

$$T_{(a)}(Z_1) - T_{(a)}(Z_2) = T_{(a)}(Z_1)[R_o(Z_2) - R_o(Z_1)]T_{(a)}(Z_2)$$

3. 32

This condition follows directly from the unitarity condition equation 2. 6 for the amplitude $\tau_{(a)}$. A formal proof of the unitarity of the three-particle amplitude will check the validity of the equation

$$T(Z_1) - T(Z_2) = T(Z_1)[R_o(Z_2) - R_o(Z_1)]T(Z_2)$$
$$T = \sum_{ab} T_{ab}$$

3. 33

It is convenient to define the partial sums over the channels as follows:

$$L_a \equiv \sum_b T_{ab}$$

$$F_b \equiv \sum_a T_{ab}$$

**3.34**

The unitarity condition to be proved is then given by

$$T(Z_1) - T(Z_2) \stackrel{?}{=}$$

$$\sum_d F_d(Z_1)[R_o(Z_2) - R_o(Z_1)]L_d(Z_2) + \sum_{ab} \bar{\delta}_{ab} F_a(Z_1)[R_o(Z_2) - R_o(Z_1)]L_b(Z_2)$$

**3.35**

The term for which the indices are the same has been separated out to make use of the two-particle unitarity condition.

This is possible by rewriting the relativistic coupled channel (Faddeev) equation using

$$L_a = T_{(a)}\left(1 - \sum_d \bar{\delta}_{ad} R_o L_d\right)$$

$$F_b = \left(1 - \sum_d \bar{\delta}_{db} F_d R_o\right) T_{(b)}$$

By direct substitution, this gives

$$T(Z_1) - T(Z_2) \stackrel{?}{=}$$

$$\sum_d \left(1 - \sum_f \bar{\delta}_{fd} F_f(Z_1) R_o(Z_1)\right) T_{(d)}(Z_1)[R_o(Z_2) - R_o(Z_1)] T_{(d)}(Z_2) \left(1 - \sum_g \bar{\delta}_{gd} R_o(Z_2) L_g(Z_2)\right) +$$

$$\sum_{ab} \bar{\delta}_{ab} F_a(Z_1)[R_o(Z_2) - R_o(Z_1)] L_b(Z_2)$$

We can now take advantage of the two-particle unitarity of the input amplitudes $T_{(d)}$ to write

$$T(Z_1) - T(Z_2) \stackrel{?}{=}$$

$$\sum_d \left(1 - \sum_f \bar{\delta}_{fd} F_f(Z_1) R_o(Z_1)\right) \left(T_{(d)}(Z_1) - T_{(d)}(Z_2)\right) \left(1 - \sum_g \bar{\delta}_{gd} R_o(Z_2) L_g(Z_2)\right) +$$

$$\sum_{ab} \bar{\delta}_{ab} F_a(Z_1)[R_o(Z_2) - R_o(Z_1)] L_b(Z_2)$$

which simplifies to

$$T(Z_1) - T(Z_2) \stackrel{?}{=} T(Z_1) - T(Z_2) +$$

$$\sum_d \left\{ -\left(1 - \sum_f \bar{\delta}_{fd} F_f(Z_1) R_o(Z_1)\right) T_{(d)}(Z_1) \sum_g \bar{\delta}_{gd} R_o(Z_2) L_g(Z_2) + \right.$$

$$\left. \sum_f \bar{\delta}_{fd} F_f(Z_1) R_o(Z_1) T_{(d)}(Z_2) \left(1 - \sum_g \bar{\delta}_{gd} R_o(Z_2) L_g(Z_2)\right) \right\} +$$

$$\sum_{ab} \bar{\delta}_{ab} F_a(Z_1) [R_o(Z_2) - R_o(Z_1)] L_b(Z_2)$$

Noting that the amplitudes satisfy the relativistic version of the Faddeev equations, this further simplifies to

$$T(Z_1) - T(Z_2) \stackrel{?}{=} T(Z_1) - T(Z_2) +$$

$$\sum_d \left\{ -F_d(Z_1) \sum_g \bar{\delta}_{gd} R_o(Z_2) L_g(Z_2) + \sum_f \bar{\delta}_{fd} F_f(Z_1) R_o(Z_1) \otimes L_d(Z_2) \right\} +$$

$$\sum_{ab} \bar{\delta}_{ab} F_a(Z_1) [R_o(Z_2) - R_o(Z_1)] L_b(Z_2) =$$

$$T(Z_1) - T(Z_2)$$

Therefore, unitarity of the two-particle input amplitudes embedded in the three-particle space implies unitarity of the full three-particle scattering amplitudes. This proof immediately generalizes to include the addition of non-separable three particle interactions of the type previously discussed, when they are unitarily implemented. This is done through the addition of a three-particle Faddeev channel $T_{(0)}$ in addition to the pair-clustering channels $T_{(a)}$. The additional channel is then summed along with the pair-clustering channels, and the proof is unchanged.

D.  Extracting Physical Observables

Typically, the scattering amplitude is decomposed into a part corresponding to the identity, and a transition amplitude. For three-particle scattering, several types of transitions are possible. In general, any of the possible boundary states, three-free or combinations of bound pair + spectator, can exist in either the initial or final state. For the general case the transition amplitude satisfies

$$< \Psi_\alpha^{(+)} : M, \underline{u} | \Psi_\beta^{(-)} : M_o, \underline{u}_o > = \delta_{\alpha\beta} < \Phi_\alpha : M, \underline{u} | \Phi_\beta : M_o, \underline{u}_o > +$$
$$2\pi i \delta^4(M\vec{u} - M_o\vec{u}_o) A_{\alpha\beta}(\Phi_\alpha | \Phi_\beta : M_o)$$

3. 36

where the other quantum numbers have been temporarily suppressed for conciseness. The four dimensional energy-momentum delta function can alternatively be expressed as

$$\delta^4(M\vec{u} - M_o\vec{u}_o) = \frac{\delta(M - M_o)}{M^3} u^0 \delta^3(\underline{u} - \underline{u}_o)$$

3. 37

The differential cross section will be written as follows using this normalization, assuming the initial state involves only two particles :

$$d\sigma = \frac{(m_{1o} m_{2o})}{[(\vec{k}_{1o} \cdot \vec{k}_{2o})^2 - m_{1o}^2 m_{2o}^2]^{\frac{1}{2}}} \prod_a \left(\frac{m_a}{\varepsilon_a} d^3 k_a\right) (2\pi)^4 \delta^4\left(\sum_a \vec{k}_a - \vec{k}_{1o} - \vec{k}_{2o}\right) |A_{fo}|^2$$

3. 38

(this form can be generalized for three-particle incident flux, though experimentally constructing such processes is usually impractical).

As is well known in scattering theory, the scattering states can be represented in terms of the boundary states using the fully interacting resolvant[6,7]

$$\left|\Psi_\alpha^{(\pm)} : M^{(\alpha)}, \underline{u}\right\rangle = \lim_{\eta \to 0} \left[\mp i\eta R_F(M^{(\alpha)} \pm i\eta)\right] \left|\Phi_\alpha : M^{(\alpha)}, \underline{u}\right\rangle$$

3. 39

where $M^{(\alpha)}$ is the appropriate invariant energy parameter for either a three-free or a pair + spectator boundary state. Thus, the scattering states can be directly extracted from the scattering amplitudes. To do this, it is convenient to define the following set of operations:

$$R_o(Z)W_{ab}(Z)R_o(Z) \equiv R_o(Z)K_{ab}(Z)R_b(Z)$$
$$\equiv R_a(Z)\tilde{K}_{ab}(Z)R_o(Z)$$
$$\equiv R_a(Z)Q_{ab}(Z)R_b(Z)$$

3. 40

These expressions will allow the physical transition amplitudes to be directly related to the calculated amplitudes. One does this by examining various ways of representing products of the full resolvant, which satisfies the equations

$$R_F(Z) = R_o(Z) - R_o(Z)\sum_{ab}T_{ab}(Z)R_o(Z)$$

3. 41

For instance, if one is interested in pair + spectator elastic scattering and re-arrangement, one would utilize the form

$$R_F(Z_1)R_F(Z_2) = R_o(Z_1)R_o(Z_2) + \sum_a [R_a(Z_1)R_a(Z_2) - R_o(Z_1)R_o(Z_2)] +$$
$$\sum_{ab} R_a(Z_1)\left\{\left[\frac{1}{Z_1 - Z_2} - R_a(Z_2)\right]Q_{ab}(Z_2) - Q_{ab}(Z_1)\left[\frac{1}{Z_1 - Z_2} - R_b(Z_1)\right]\right\}R_b(Z_2)$$

3. 42

Typically, when going on-shell in the parameter Z, final invariant energy conservation results from the usual relation

$$\frac{1}{M - (M_o \pm i\eta)} \xrightarrow{\eta \to 0^+} \frac{\wp}{M - M_o} \pm i\pi\delta(M - M_o)$$

3. 43

where $\wp$ represents the Cauchy principle part.

A physical boundary state that includes a bound state as one of its constituent systems will have a "primary singularity" [see references 13,6,7] in the off-shell parameter Z corresponding to that bound state. In two-particle scattering theory, these primary singularities can be used to extract the bound state from the scattering amplitude using

$$\underset{\varsigma \to \varepsilon_{\mu_{(a)}}}{Lim}\left[(\varsigma-\mu_{(a)})\tau_{(a)}(M_{(a)},\hat{q}_{(a)}|M'_{(a)},\hat{q}'_{(a)};\varsigma)\right]=$$

$$(M_{(a)}-\mu_{(a)})\psi_{(a)}(M_{(a)},\mu_{(a)},l,l_z)Y_l^{l_z}(\hat{q}_{(a)})Y_l^{l_z*}(\hat{q}'_{(a)})\psi_{(a)}^*(M'_{(a)},\mu_{(a)},l,l_z)(M'_{(a)}-\mu_{(a)})$$

<div align="right">3. 44</div>

One utilizes this singular behavior in the three-particle scattering to extract the physical amplitudes from the scattering amplitudes $W_{ab}$ or $T_{ab}$. Using the example given above, one can calculate amplitudes for pair + spectator elastic scattering and rearrangement by using

$$\underset{Z \to (\varepsilon_{bo}+\varepsilon_{\mu_{(b)o}})\pm i0}{Lim}\left[(Z-(\varepsilon_a+\varepsilon_{\mu_{(a)}}))(Z-(\varepsilon_{bo}+\varepsilon_{\mu_{(b)o}}))\right]\otimes$$

$$u^0\delta^3(\underline{u}-\underline{u}_o)\frac{W_{ab}(M,\underline{u}_{(a)},\hat{q}_{(a)}|M_o,\underline{u}_{(b)o},\hat{q}_{o(b)};Z)}{\sqrt{\rho_{(a)}^{(3)}(M,u_{(a)}^0)}\sqrt{\rho_{(b)}^{(3)}(M_o,u_{(b)o}^0)}}=$$

$$(M-(\varepsilon_a+\varepsilon_{\mu_{(a)}}))\frac{\psi_{(a)}(M_{(a)},\mu_{(a)},l_{(a)},l_{z(a)})}{\sqrt{\rho_{(a)}^{(2)}(M_{(a)},m_{a+},m_{a-})}}Y_{l_{(a)}}^{l_{z(a)}}(\hat{q}_{(a)})\otimes$$

$$Q_{ab}^{(\pm)}(\underline{k}_a,m_a;\psi_{(a)}(\underline{p}_{(a)},\mu_{(a)},l_{(a)},l_{z(a)})|\underline{k}_{bo},m_b;\psi_{(b)}(\underline{p}_{(b)o},\mu_{(b)},l_{(b)o},l_{z(b)o}):(\varepsilon_{bo}+\varepsilon_{\mu_{(b)o}}))\otimes$$

$$Y_{l_{(b)}}^{l_{z(b)}*}(\hat{q}_{o(b)})\frac{\psi_{(b)}^*(M_{(b)o},\mu_{(b)},l_{(b)o},l_{z(b)o})}{\sqrt{\rho_{(b)}^{(2)}(M_{(b)o},m_{b+},m_{b-})}}(M_o-(\varepsilon_{bo}+\varepsilon_{\mu_{(b)o}}))$$

<div align="right">3. 45</div>

where the parameters $M_{(a)}$ are the invariant pair energy parameters as expressed in equations 2. 1 and 3. 5. This extracts a pole from both the initial and final states. Note that if the physical system consists of particle a scattering from a confined pair a+ and a-, the only physically relevant amplitude will be $Q_{aa}$. The other physical amplitudes can be extracted in a similar manner. We are therefore able to express the physical amplitudes directly in terms of the amplitudes defined in equation 3. 40.

<div align="center">Elastic and rearrangement scattering</div>

$$\left\langle\Phi_{(a)}:\underline{k}_a,m_a;\psi_{(a)}(\underline{p}_{(a)},\mu_{(a)},l_{(a)},l_{z(a)})\left|Q_{ab}^{(+)}(M_o)\right|\Phi_{(b)}:\underline{k}_b,m_b;\psi_{(b)}(\underline{p}_{(b)o},\mu_{(b)},l_{(b)o},l_{z(b)o})\right\rangle\otimes$$

$$\delta(M-M_o)=$$

$$-A_{ab}^{(+)}(\underline{k}_a,m_a;\psi_{(a)}(\underline{p}_{(a)},\mu_{(a)},l_{(a)},l_{z(a)})|\underline{k}_{bo},m_b;\psi_{(b)}(\underline{p}_{(b)o},\mu_{(b)},l_{(b)o},l_{z(b)o}):M_o)\otimes$$

$$\delta^4(\vec{k}_a+\vec{p}_{(a)}-\vec{k}_{bo}-\vec{p}_{(b)o})$$

<div align="right">3. 46</div>

where $M_o = \varepsilon_{bo} + \varepsilon_{\mu_{(b)o}}$ and $M = \varepsilon_a + \varepsilon_{\mu_{(a)}}$.

### Breakup

$$\left\langle \Phi_{(0)} : \underline{k}_1, m_1; \underline{k}_2, m_2; \underline{k}_3, m_3 \left| \sum_a K_{ab}^{(+)}(M_o) \right| \Phi_{(b)} : \underline{k}_b, m_b; \psi_{(b)}(\underline{p}_{(b)o}, \mu_{(b)}, l_{(b)o}, l_{z(b)o}) \right\rangle \otimes$$
$$\delta(M - M_o) =$$
$$- A_{0b}^{(+)}(\underline{k}_1, m_1; \underline{k}_2, m_2; \underline{k}_3, m_3 \mid \underline{k}_{bo}, m_b; \psi_{(b)}(\underline{p}_{(b)o}, \mu_{(b)}, l_{(b)o}, l_{z(b)o}) : M_o) \otimes$$
$$\delta^4(\vec{k}_1 + \vec{k}_2 + \vec{k}_3 - \vec{k}_{bo} - \vec{p}_{(b)o})$$

3. 47

where $M_o = \varepsilon_{bo} + \varepsilon_{\mu_{(b)o}}$.

### Coalescence

$$\left\langle \Phi_{(a)} : \underline{k}_a, m_a; \psi_{(a)}(\underline{p}_{(a)}, \mu_{(a)}, l_{(a)}, l_{z(a)}) \left| \sum_b \tilde{K}_{ab}^{(+)}(M) \right| \Phi_{(0)} : \underline{k}_{1o}, m_1; \underline{k}_{2o}, m_2; \underline{k}_{3o}, m_3 \right\rangle \otimes$$
$$\delta(M - M_o) =$$
$$- A_{a0}^{(+)}(\underline{k}_a, m_a; \psi_{(a)}(\underline{p}_{(a)}, \mu_{(a)}, l_{(a)}, l_{z(a)}) \mid \underline{k}_{1o}, m_1; \underline{k}_{2o}, m_2; \underline{k}_{3o}, m_3 : M) \otimes$$
$$\delta^4(\vec{k}_a + \vec{p}_{(a)} - \vec{k}_{1o} - \vec{k}_{2o} - \vec{k}_{3o})$$

3. 48

where $M = \varepsilon_a + \varepsilon_{\mu_{(a)}}$.

### Three to three scattering

$$\left\langle \Phi_{(0)} : \underline{k}_1, m_1; \underline{k}_2, m_2; \underline{k}_3, m_3 \left| \sum_b T_{ab}^{(+)}(M_o) \right| \Phi_{(0)} : \underline{k}_{1o}, m_1; \underline{k}_{2o}, m_2; \underline{k}_{3o}, m_3 \right\rangle \otimes$$
$$\delta(M - M_o) =$$
$$- A_{00}^{(+)}(\underline{k}_1, m_1; \underline{k}_2, m_2; \underline{k}_3, m_3 \mid \underline{k}_{1o}, m_1; \underline{k}_{2o}, m_2; \underline{k}_{3o}, m_3 : M_o) \otimes$$
$$\delta^4(\vec{k}_1 + \vec{k}_2 + \vec{k}_3 - \vec{k}_{1o} - \vec{k}_{2o} - \vec{k}_{3o})$$

3. 49

where $M_o$ is the on-shell invariant energy of the three-particle system. The physical (non-singular) amplitudes $A_{\alpha\beta}$ can then be directly substituted into equation 3. 38 to calculate the cross section for a particular type of process.

## IV. Limiting Behavior

### A. Cluster Decomposability

Perhaps the most difficult of the physical requirements to incorporate into the formalism was its cluster decomposition properties. Physically, one would not expect the kinematic behavior of a relativistic electron in another galaxy to modify the dynamical spectrum of hydrogen atoms or positronium here on earth. In a scattering theory, the complicated non-linear energy-momentum dispersion relations of Lorentz covariant transformations make the incorporation of this straightforward expectation non-trivial. For our purposes, we will consider a set of cluster decomposed systems to describe physical situations in which there is no "quantum entanglement" of the spectating clusters with the dynamical cluster. In a sense, the spectating clusters are then considered to be classically disjoint. This is, of course, the usual boundary state assumption in scattering theory.

The form of the relativistic coupled equations for the amplitudes 3.19 decouples when the system cluster decomposes as described above. If the spectating cluster or particle has label $a$ for the decomposed system, there are no interactions or entanglements between that particle and either of the other dynamical particles, which implies that $T_{(d)}=0$ for $d \neq a$. Thus, the dynamical equation becomes

$$W_{ab}(Z) = -\bar{\delta}_{ab} T_{(a)}(Z) R_o(Z) T_{(b)}(Z) - \sum_d \bar{\delta}_{ad} T_{(a)}(Z) R_o(Z) W_{db}(Z) \Rightarrow 0$$

$$T_{ab}(Z) = \delta_{ab} T_{(a)}(Z) + W_{ab}(Z) \Rightarrow \delta_{ab} T_{(a)}(Z)$$

4.1

This means that the form of the embedding of the two-particle amplitudes into the three-particle space will determine the clustering properties of the formalism.

One's initial tendency to utilize three-momentum conservation to constrain the kinematics (which works well in non-relativistic quantum scattering) encounters conceptual difficulties when the system is relativistic. To see this, first note that the velocity of the Lorentz frame of a system with momentum $\underline{P}$ and energy E is given by

$$\underline{v} = \frac{\underline{P}}{E}.$$

In formulations of scattering theory in which intermediate states are off-diagonal in the energy E', one sees immediately that if momentum is conserved $(\underline{P} = \underline{P}')$, then the Lorentz frames of the components of the system are different:

$$\underline{v} = \frac{\underline{P}}{E} \neq \frac{\underline{P}'}{E'} = \underline{v}'$$

**4. 2**

Thus, in order to formulate the clustering property, off-diagonal momentum conservation was relaxed.

The form that satisfies the authors' criteria for cluster decomposability is reproduced below for the discussion that follows:

$$\langle \underline{k}_a \underline{k}_{a+} \underline{k}_{a-} | T_{(a)}(Z) | \underline{k}'_a \underline{k}'_{a+} \underline{k}'_{a-} \rangle = (u^0) \delta^3(\underline{u} - \underline{u}')(u^0_{(a)}) \delta^3(\underline{u}_{(a)} - \underline{u}'_{(a)}) \frac{\tau_{(a)}(\omega_{(a)} \hat{q}_{(a)} | \omega'_{(a)} \hat{q}'_{(a)}; \zeta_{(a)})}{\sqrt{[\rho^{(3)}_{(a)}(M, u^0_{(a)}) \rho^{(3)}_{(a)}(M', u^0_{(a)})]}}$$

Several features of this form should be noted. The Lorentz invariance relative to the overall system's reference frame requires conservation of the overall velocity $\underline{u}$, which in a product with an invariant energy diagonalizing delta function from the resolvant [see equation 3. 43] will give full four-momentum conservation for the system. Since momentum conservation has been relaxed off-diagonal, the conservation of the spectator's momentum utilized in Faddeev's non-relativistic approach is inadequate to properly constrain the kinematics of the dynamical pair. The inclusion of the conservation of pair velocity $\underline{u}_{(a)}$ does this in a Lorentz invariant way, and it similarly becomes equivalent to pair four-momentum conservation if the invariant energy off-diagonal parameterization of the pair is done properly.

If the energy of the spectating cluster does not enter into the dynamical equations parametrically, then the kinematics of the spectator would influence the dynamics of the interacting (dynamical) subsystem, which would violate a primary tenet of cluster decomposability. Typically, the intermediate integrals would mix up the kinematics of the spectator with the dynamics of the pair in such a way as to alter bound state poles in the off-shell parameter, for instance. The use of the off-shell parameter $\zeta_{(a)}$ and the off-diagonal parameter $\omega_{(a)}$ prevent these complications.

$$e_a(M_o u^0_{(a)}) \equiv u^0_{(a)} \sqrt{m_a^2 + M_o^2 u^2_{(a)}} - M_o u^2_{(a)}$$

$$\zeta_{(a)} = \frac{Z - e_a}{u^0_{(a)}}, \omega_{(a)} = \frac{M - e_a}{u^0_{(a)}}$$

The off-shell parameter has only parametric dependency on the relativistic kinematics of the spectator and the on-shell invariant energy $M_o$, and since the off-diagonal parameter has this same dependency upon the spectator kinematics, the dynamical spectrum of the pair remains unchanged; it is only expressed using total invariant energy kinematics. The factor $u_{(a)}^0$ in the denominators of these parameters is just the Lorentz factor which appropriately expresses invariant pair energies in the three-particle center of momentum frame. If the two-particle input amplitude is embedded in the three-particle space using the intermediate pair energy $M_{(a)}$ derivable from the momenta, then the intermediate integrals would entail complicated dependencies on the kinematics of the spectating cluster which would entangle its independent motions in non-trivial ways.

The velocity conserving form used by the authors is formally similar to the point form of relativistic dynamics of Dirac[23]. By virtue of the commutation relation $[K_j, P_k] = i\delta_{jk} H$ between these generators in the Poincare group, interactions present in the Hamiltonian generator H will also manifest in either the boost generator $\underline{K}$ (instant form) or the displacement generator $\underline{P}$ (point form). The use of point form formulations was shown to be equivalent to the instant form and the light-front form by Sokolov[24], but the point form has the convenience of manifest Lorentz invariance. The point form maintains a well defined Lorentz frame, but in terms of the boundary states' momenta translations are generated by an off-diagonal operator, and if represented in terms of this operator can appear to be non-local. However, represented in terms of the fully interacting operators, all interactions remain local. Similarly, the instant form maintains spatial relationships represented in terms of either set of operators; however, the Lorentz frame becomes ill defined off-diagonal. Sokolov[25] used the point form to achieve cluster separability for a system of directly interacting particles. However, in our approach the authors make no presumption about the microscopic form of or need for interaction potentials, only that the boundary states be eigenstates with well defined energies and momenta. Brodsky and Ji[26] showed that in the zero binding limit, a relativistic wave function in the light-front form using a simple field theoretic model could be shown to factor into two separate wave functions. The authors wish to emphasize that the cluster decomposibility of our formulation is not model dependent. A finite set of possible forms of relativistic dynamics can be constructed in terms of the generators of the Poincare group; the interested reader is referred to the literature for further discussion.[27]

The clustering properties make the usefulness of this formulation particularly apparent. In physical problems which involve boundary states which are bound states or confined states of elementary particles, the appropriate exterior channel of the coupled equations necessarily must contain a primary singularity with the proper kinematics, and thus must be associated with a cluster decomposed channel. Such restrictions on the structure of the form of the amplitudes are particularly useful for atomic and nuclear scatterings, and boundary state hadronization of strongly interacting particles.

This last point, although discussed some time ago[12], is often not immediately apparent to those not familiar with relativistic Faddeev type equations. Suppose we wish to construct a theory based upon confined constituents (e.g. quarks and gluons) which, by hypothesis, can never be directly observed as asymptotic particles. The basic point is that in a formalism of the type presented here, cluster decomposability allows clusters of arbitrary complexity to appear as boundary states. If, by hypothesis, only clusters containing "confined particles" (and when the theory is extended, antiparticles and quanta) and not the "particles" themselves can appear asymptotically, then the actual dynamical equations from which these clusters are extracted remain unchanged. As an example consider the confined system discussed immediately after equation 3.45. Of course, to go from our formalism as presented here to phenomenology or a theory of confinement capable of meeting well know types of observations will require extension of the formalism to show how to construct *unitary* models and/or theories of two particle sub-amplitudes which can be used as input.

### B. Non-Relativistic Limit

We now show that our relativistic formalism reduces unambiguously to the non-relativistic quantum dynamics of a finite number of particles and their bound states. The complexity of the relativistic velocity transformations, which we found we had to use in order to preserve both Lorentz invariance and cluster decomposability, makes it by no means obvious what form that non-relativistic theory will take. Fortunately for us, we can show that our theory does reduce unambiguously to the non-relativistic Faddeev equations. Since we allow (but need not require) our model to be specified fully off shell and off diagonal, all we then need to construct any corresponding non-relativistic ``potential'' is to require that our non-relativistic limit is time-reversal invariant. Then the Lippmann-Schwinger equation can be used to construct the potential[14]. But the Faddeev equations themselves are more flexible than potential models, and can provide us with additional insights (see Conclusion). Because of the importance of the result, we try to omit no crucial step in the derivation that follows. Our strategy is to first discuss the (by hypothesis, *unitary*) two-body input, and then embed it in the three-body space.

We start with the two-particle kinematics in the two-body space. The momentum $\underline{q}_{(a)}$ is just the (boost-invariant) momentum of either particle in the 2-particle center of momentum frame. Relativistically $\underline{q}_{(a)}$ satisfies the usual energy-momentum relationship in the pair center of momentum system given in equation 2.2. Using the symbol "$\Rightarrow$" for the non-relativistic limit, the non-relativistic form of the momentum $\underline{q}_{(a)}$ can be shown to be given by

$$\underline{q}_{(a)} \Rightarrow \frac{m_{a+}\underline{k}_{a-} - m_{a-}\underline{k}_{a+}}{m_{a+} + m_{a-}}$$

4. 3

which can readily be seen to be the usual form using non-relativistic kinematics. Substitution into the equation 2.1 for $M_{(a)}$ gives the usual non-relativistic form of the energy of the pair

$$M_{(a)} \Rightarrow m_{a+} + m_{a-} + \frac{q_{(a)}^2}{2m_a^{reduced}}$$

4. 4

where the reduced mass satisfies the usual relationship

$$m_a^{reduced} \equiv \frac{m_{a+}m_{a-}}{m_{a+} + m_{a-}}$$

The non-relativistic limit of the 2-particle scattering amplitude conserves momentum, and corresponds to the usual non-relativistic amplitudes as follows:

$$\langle \underline{k}_{a+}\underline{k}_{a-} | t_{(a)}(m_{a+} + m_{a-} + z_{NR}) | \underline{k}_{a+0}\underline{k}_{a-0}\rangle \Rightarrow \delta^3(\underline{k}_{a+} + \underline{k}_{a-})\langle \underline{q}_{(a)} | t_{(a)}^{NR}(z_{NR}) | \underline{q}_{(a)0}\rangle$$

$$\equiv \delta^3(\underline{k}_{a+} + \underline{k}_{a-}) \frac{\tau_{(a)}^{NR}(E_{(a)}, \hat{q}_{(a)} | E_{(a)0}, \hat{q}_{(a)0}; z_{NR})}{\sqrt{m_a^{reduced} q_{(a)}}\sqrt{m_a^{reduced} q_{(a)0}}}$$

4. 5

or, by direct comparison

$$\tau_{(a)}\left(m_{a+} + m_{a-} + E_{(a)}, \hat{q}_{(a)} | m_{a+} + m_{a-} + E_{(a)0}, \hat{q}_{(a)0}; m_{a+} + m_{a-} + z_{NR}\right) \Rightarrow$$
$$\tau_{(a)}^{NR}\left(E_{(a)}, \hat{q}_{(a)} | E_{(a)0}, \hat{q}_{(a)0}; z_{NR}\right)$$

4. 6

When we embed this interacting pair in the center of momentum system of the spectator plus pair with invariant mass $M_{(a)}$, the spectator will have 3-momentum $-(m_{a+} + m_{a-})\underline{u}_a$ and the interacting pair $(m_{a+} + m_{a-})\underline{u}_a$. In the non-relativistic limit

$$\underline{p}_a \equiv \frac{m_a(\underline{k}_{a+} + \underline{k}_{a-}) - (m_{a+} + m_{a-})\underline{k}_a}{m_a + m_{a+} + m_{a-}} \Rightarrow -\underline{k}_a$$

4.7

and the invariant mass $M$ of the three particle system goes to

$$M \Rightarrow m_a + m_{a+} + m_{a-} + \frac{p_a^2}{2n_a} + \frac{q_a^2}{2m_a^{reduced}}$$

4.8

where

$$\frac{1}{n_a} = \frac{1}{m_a} + \frac{1}{m_{a+} + m_{a-}}$$

This is precisely the form expected for the non-relativistic kinematics.

In the three-particle space the non-relativistic limit of the Jacobian $\rho_{(a)}^{(3)}$ from equation 3.4 takes the form

$$\rho_{(a)}^{(3)} \Rightarrow (m_1 + m_2 + m_3)^3 (m_{a+} + m_{a-})^3 \frac{m_{a+} m_{a-}}{m_{a+} + m_{a-}} |\underline{q}_{(a)}|$$

4.9

The non-relativistic limit of the off-shell parameter $\zeta_{(a)}$ obtained by direct substitution of the form $Z = m_1 + m_2 + m_3 + z$, and using the non-relativistic form for the parameter $e_a$, is given by

$$\zeta_{(a)} \Rightarrow m_{a+} + m_{a-} + z - \frac{k_a^2}{2n_a}$$

4.10

The non-relativistic form of the off-diagonal parameter $\omega_{(a)}$ is given in terms of the non-relativistic pair energy $E_{(a)}$ by the expression

$$\omega_{(a)} \Rightarrow m_{a+} + m_{a-} + E_{(a)}$$

4. 11

Substituting these limiting forms into the amplitude connection given in equation 3. 12, the non-relativistic limit of the embedded cluster decomposable amplitude directly becomes the Faddeev input amplitude.

$$\tau_{(a)}\left(\omega_{(a)}\hat{q}_{(a)} \mid \omega'_{(a)}\hat{q}'_{(a)};\zeta_{(a)}\right) \Rightarrow \tau_{(a)}^{NR}\left(E_{(a)}\hat{q}_{(a)} \mid E'_{(a)}\hat{q}'_{(a)}; z - \frac{k_a^2}{2n_a}\right)$$

4. 12

From the $W_{ab}$ equation 3. 21 one can take the non-relativistic limit of the $F_{(ad)}$ factor from equation 3. 25

$$F_{(ad)}\left(u_{(a)}, u'_{(d)}, \hat{u}_{(a)} \cdot \hat{u}'_{(d)}\right) \approx$$

$$\left[\frac{\varepsilon_a^{(a)}\varepsilon_d^{(d)}}{q'_{(a)}q'_{(d)}}\right]^{\frac{1}{2}} \frac{M'^2_{(a)}M'^2_{(d)}}{\varepsilon'_d\varepsilon_a^{(d)}\varepsilon_{d-}^{(d)}u_{(a)}^0} \Rightarrow \left[\frac{m_a m_d}{q'_{(a)}q'_{(d)}}\right]^{\frac{1}{2}} \frac{(m_{a+} + m_{a-})^2(m_{d+} + m_{d-})^2}{m_a m_d m_{d-}}$$

4. 13

This can be substituted into the equation for the amplitudes $W_{ab}$

$$\frac{W_{ab}\left(M\underline{u}_{(a)}, \hat{q}_{(a)} \mid M_o\underline{u}_{o(b)}, \hat{q}_{o(b)}; m_1 + m_2 + m_3 + z\right)}{\sqrt{\rho_{(a)}^{(3)}(M, u_{(a)}^0)}\sqrt{\rho_{(b)}^{(3)}(M_o, u_{o(b)}^0)}} \Rightarrow W_{ab}^{D-NR}\left(\underline{p}_{(a)}, \underline{q}_{(a)} \mid \underline{p}_{o(b)}, \underline{q}_{o(b)}; z\right) +$$

$$-\sum_d \bar{\delta}_{ad} \int \frac{(m_{d+} + m_{d-})^3 d^3 u'_{(d)}}{m_a m_d m_{d-}} \frac{m_a(m_{a+} + m_{a-})}{\sqrt{q_{(a)}q'_{(a)}}} \tau_{(a)}^{NR}\left(E_{(a)}\hat{q}_{(a)} \mid E'_{(a)}\hat{q}'_{(a)}; z - \frac{k_a^2}{2n_a}\right) \otimes$$

$$\frac{1}{M'^{(d)} - (m_1 + m_2 + m_3 + z)} W_{db}\left(M'\underline{u}'_{(d)}, \hat{q}'_{(d)} \mid M_o\underline{u}_{o(b)}, \hat{q}_{o(b)}; m_1 + m_2 + m_3 + z\right)$$

4. 14

The kinematic factors relating non-relativistic energy amplitudes to momentum amplitudes are given by

$$\tau_{(a)}^{NR}\left(E_{(a)}\hat{q}_{(a)} \mid E'_{(a)}\hat{q}'_{(a)}; z - \frac{k_a^2}{2n_a}\right)\left(\frac{m_{a+} + m_{a-}}{m_{a+}m_{a-}\sqrt{q_{(a)}q'_{(a)}}}\right) = \tau_{(a)}^{NR}\left(\underline{q}_{(a)} \mid \underline{q}'_{(a)}; z - \frac{p_a^2}{2n_a}\right)$$

Hence

$$\frac{W_{ab}\left(M \underline{u}_{(a)}, \hat{q}_{(a)} | M_o \underline{u}_{o(b)}, \hat{q}_{o(b)}; m_1 + m_2 + m_3 + z\right)}{\sqrt{\rho^{(3)}_{(a)}(M, u^0_{(a)})}\sqrt{\rho^{(3)}_{(b)}(M_o, u^0_{o(b)})}} \Rightarrow$$

$$W^{NR}_{ab}\left(\underline{p}_{(a)}, \underline{q}_{(a)} | \underline{p}_{o(b)}, \underline{q}_{o(b)}; z\right) = W^{D-NR}_{ab}\left(\underline{p}_{(a)}, \underline{q}_{(a)} | \underline{p}_{o(b)}, \underline{q}_{o(b)}; z\right) +$$

$$- \sum_d \bar{\delta}_{ad} \int d^3 p'_{(d)} \tau^{NR}_{(a)}\left(\underline{q}_{(a)} | \underline{q}'_{(a)}; z - \frac{p_a^2}{2n_a}\right) \frac{1}{E'-z} W^{NR}_{db}\left(\underline{p}'_{(d)}, \underline{q}'_{(d)} | \underline{p}_{o(b)}, \underline{q}_{o(b)}; z\right)$$

4. 15

Using the form of the connection between the momenta $\underline{p}'_{(d)}, \underline{q}'_{(d)}$ and $\underline{p}'_{(a)}, \underline{q}'_{(a)}$, i.e.

$$\underline{p}'_{(a)} = -\left(\underline{q}'_{(d)} + \frac{m_a}{m_a + m_d}\underline{p}'_{(d)}\right)$$

4. 16

this can directly be written in the form given by Faddeev[13]

$$W^{NR}_{ab}\left(\underline{p}_{(a)}, \underline{q}_{(a)} | \underline{p}_{o(b)}, \underline{q}_{o(b)}; z\right) = W^{D-NR}_{ab}\left(\underline{p}_{(a)}, \underline{q}_{(a)} | \underline{p}_{o(b)}, \underline{q}_{o(b)}; z\right) +$$

$$- \sum_d \bar{\delta}_{ad} \int d^3 p'_{(d)} d^3 q'_{(d)} \delta^3(\underline{p}_{(a)} - \underline{p}'_{(a)}) \tau^{NR}_{(a)}\left(\underline{q}_{(a)} | \underline{q}'_{(a)}; z - \frac{p_a^2}{2n_a}\right) \otimes$$

$$\frac{1}{E'-z} W^{NR}_{db}\left(\underline{p}'_{(d)}, \underline{q}'_{(d)} | \underline{p}_{o(b)}, \underline{q}_{o(b)}; z\right)$$

4. 17

which is simply the connected non-relativistic Faddeev equation. Q.E.D.

## Conclusion

We have now articulated a general, Lorentz invariant, cluster decomposable, unitary and calculable quantum mechanical scattering theory for any finite number of distinct particles representable as free particle boundary states. Although the explicit theory presented here is for three scalar particles, the extension to binary clusters containing any finite number of particles is immediately available by recursion. This recursive approach to the proof of cluster decomposability is also needed for that proof in a perturbative relativistic quantum field theory for the same boundary states, as can be seen in the general discussion of clustering given by Weinberg[28] (see e.g. Ch. 4). We emphasize that our approach succeeds where other attempts to combine Lorentz invariance, unitarity and cluster decomposability have failed because we use the appropriate relativistic extension of the Faddeev theory. This extension keeps the dynamical (interacting) cluster and the motion of the spectator in the *same* Lorentz 4-velocity reference frame. Then the (necessarily) off-shell description of the internal dynamics of the interacting cluster only depends parametrically on the kinematics of the spectator, in a manner that is analogous to the method by which non-relativistic Faddeev theory achieves cluster decomposability. The price we have to pay is to carry out the algebra of the Lorentz velocity transformations when we go from one cluster to another and calculate the (admittedly complicated) Jacobians. That we have done so explicitly makes quantitative application of the theory straightforward. The restriction to scalar particles has already been removed in previous work[8] by using the convenient properties of the helicity representation of angular momentum. Clearly our formalism opens up a host of phenomenological applications to quantitative exploration. We emphasize that these applications will be facilitated by the fact that our input variables are *physical parameters* such as masses of isolated systems, conserved charges, and angular momenta. As such they can be specified by appropriate experimental observables.

The fact that the formalism reduces directly and unambiguously to the general Faddeev theory in the non-relativistic limit --- which we emphasize was *not* built in as a requirement --- yields an added bonus. Provided the relativistic form for the input two-body amplitudes is time-reversal invariant, and derived from some physical theory, this non-relativistic limit allows us to calculate a form of corresponding non-relativistic ``potentials'', giving us a new way of meeting an old problem.

We emphasize that this formulation is NOT meant to be a fixed number approximation to physical scattering processes, but rather a non-perturbative mechanism for including unitary processes in an overall unitary calculation including a finite number of particles in the initial and final states. For the formalism to incorporate any finite number of elementary particles and quanta, we must take several more steps. The first is to introduce the particle-antiparticle symmetry properties found in nature into the two particle input in such a way that a unitary particle-particle amplitude leads unambiguously to a *unitary* particle-antiparticle amplitude. Then the FLN[4] proof of unitarity will carry over to the expanded system. A paper showing how to do this in such a way to extend these amplitudes to include particle-antiparticle scattering amplitudes and the transition from one particle-antiparticle pair to another has been prepared[29].

The second step is to show that some particle-particle and particle-antiparticle scattering channels contain coherent amplitudes with singularities corresponding to appropriate quanta. A second paper[30] will show that in a model which has quantum exchange between particle pairs, if the amplitudes are properly embedded in a three particle space one can extract quantum-particle scattering amplitudes from the resulting three-particle equations. This allows the construction of a particle-antiparticle scattering amplitude in turn, by using the identification and replacement operation presented in the first paper, and can be used to describe the process particle-antiparticle ↔ quantum-quantum (annihilation). Similarly, starting from a two particle boundary state in this three-particle space, one should also be able to extract a two-particle ↔ two-particle + particle-antiparticle process from it (pair creation). Such a finite (but not fixed) number scattering process would be associated with amplitudes generated by the formalism presented here as exemplified through a diagram such as

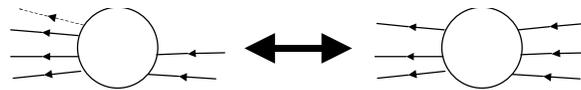

In a cluster decomposable formalism, such amplitudes should generate unitary scattering processes below appropriate scattering thresholds, and incorporate new thresholds in a straightforward manner. Our research on these problems has given us confidence that any expected behaviors can be met successfully. Results will be presented in forthcoming papers now in preparation.

**Acknowledgements**:

J.V.L. and P.K. would like to acknowledge the hospitality of the University of Dar Es Salaam Department of Physics during the period 1985-87 when a portion of this work was completed. J.V.L. would like to acknowledge the hospitality of the Stanford Linear Accelerator Center during several periods of working out the details of these calculations. The authors wish to thank Alex Markevich and Walter Lamb for careful reading and discussion of the manuscript. H.P.N. wishes to thank the director of SLAC for arranging the visitor program which enabled J.V.L. and H.P.N. to work out the final details of this paper.